\documentclass[pra,twocolumn,groupedaddress,floats,showpacs,citeautoscript]{revtex4}

\usepackage{graphicx}
\usepackage{amsmath}
\usepackage{amssymb}
\usepackage{times}
\usepackage{colordvi}

\usepackage{bm}

\newcommand* {\ee}{\ensuremath{\mathrm{e}}}

\begin{document}

\title{Sign of tunnel coupling in barrier-separated Bose-Einstein condensates and stability of double-ring systems}

\author{J. Brand}
\affiliation{Centre for Theoretical Chemistry and Physics and Institute of Natural Sciences,
Massey University (Albany Campus), Private Bag 102~904, North Shore MSC, Auckland
0745, New Zealand}

\author{T.~J. Haigh}
\affiliation{Institute of Fundamental Sciences, Massey University (Manawatu Campus),
Private Bag 11~222, Palmerston North 4442, New Zealand}
\affiliation{ARC Centre of Excellence for Quantum-Atom Optics, School of
Physical Sciences, University of Queensland, Brisbane, QLD 4072, Australia}
\altaffiliation[Present address.]{}

\author{U. Z\"ulicke}
\affiliation{Institute of Fundamental Sciences, Massey University (Manawatu Campus),
Private Bag 11~222, Palmerston North 4442, New Zealand}
\affiliation{Centre for Theoretical Chemistry and Physics, Massey University (Albany Campus),
Private Bag 102~904, North Shore MSC, Auckland 0745, New Zealand}

\date{Draft of \today.}

\begin{abstract}
We revisit recent claims about the instability of non-rotating tunnel coupled annular
Bose-Einstein condensates leading to the emergence of angular-momentum Josephson
oscillation [Phys. Rev. Lett. {\bf 98}, 050401 (2007)]. It was predicted that all stationary states
with uniform density become unstable in certain parameter regimes. By careful analysis, we
arrive at a different conclusion. We show that there is a stable non-rotating and uniform
ground state for any value of the tunnel coupling and repulsive interactions. The instability of
an excited state with $\pi$ phase difference between the condensates can be interpreted in
terms of the familiar snake instability. We further discuss the sign of the tunnel coupling
through a separating barrier, which carries significance for the nature of the stationary states.
It is found to always be negative for physical reasons. 

\end{abstract}


\maketitle

\section{Introduction}

Bose-Einstein condensates (BECs) located in different minima of an external potential created 
by magnetic or light forces and coupled by tunneling through a potential barrier have been the 
host to many exciting developments and discoveries in recent years~\cite{oberthal:jphysb:07}. 
Phenomena explored include analogs of the Josephson effect in double or multiple
quantum-well structures~\cite{albiez:prl:05,shin:prl:05,levy:nat:07}, gap solitons of repulsive
BECs~\cite{eier:prl:04}, and quantum phase transitions~\cite{greiner:nat:02}. Often these
systems are modeled by considering just one mode per potential minimum and their linear
coupling provided by tunnling through a separating barrier. These simplified models, which are
usually labelled as two-mode or multiple-mode models, variants of the Bose-Hubbard model,
or the discrete nonlinear Schr{\"o}dinger equation, are tailored to describe certain properties or
aspects of the dynamics of the many-body system under investigation. Although there is an
abundance of literature on such models~\cite{milburn:pra:97,smerzi:prl:97,sols:pra:98,roustekoski:pra:98,spekkens:pra:99,masiello:pra:07,giovanazzi:njp:08,javanainen,ananikian:pra:06,pitaevskii}, 
there still appear to exist misconceptions about the nature of the effective model parameters, 
especially the sign of the tunnel coupling, as only few works attempt to calculate such
parameters based on a more complete theoretical
treatment~\cite{ananikian:pra:06,spekkens:pra:99,giovanazzi:njp:08}.

The sign of the tunnel coupling bears special significance in systems where the tunneling 
appears over an extended (at least one-dimensional) region of space. Such systems have 
recently been analysed by Bouchoule~\cite{bou:epjd:05} and Kaurov and
Kuklov~\cite{kau:pra:06}, who studied two parallel tunnel-coupled cigar-shaped BECs. In 
another recent work, Lesanovsky and von Klitzing studied the stability of tunnel-coupled 
annular BECs~\cite{vKlitz:prl:07}. The latter paper points to an interesting dynamical instability 
leading to the spontaneous formation of angular-momentum fluctuations. We will show further 
below that the sign on the tunnel coupling bears consequences on the nature and stability of 
the stationary states found in the mean-field treatment of tunnel-coupled BECs. Specifically, 
we find that the system studied by Lesanovsky and von Klitzing has a stable ground state for 
any value of the tunnel coupling and repulsive interactions. The instability of an {\em excited\/}
state with $\pi$ phase difference between the condensates can be interpreted in terms of the
familiar snake instability~\cite{brand:pra:02}. The ground state of a rotating co-planar double-ring system is discussed in Ref.~\cite{brand09}.

We examine the stationary states of double-ring BECs in Sec.~\ref{sec:neg}. A careful
analysis of the sign of the tunnel coupling used in effective models for BECs in double-well
traps follows in Sec.~\ref{sec:sign}. Conclusions are presented in Sec.~\ref{sec:concl}.

\section{Stability of stationary states in double-ring BECs}
\label{sec:neg}

The calculation performed in Ref.~\cite{vKlitz:prl:07} starts from a number of generally
reasonable assumptions. Under the condition that radial excitations of the vertically stacked
annular BECs are suppressed by the trapping potentials and the only  mechanism for coupling
the two systems is via tunneling through a potential barrier, the Gross-Pitaevskii equation
for the two-mode spinor wave function $(\chi_{\text{u}}, \chi_{\text{d}})$ specialises to
\begin{equation}\label{chiEqs}
i \partial_\tau \chi_{\text{u/d}} = -\partial^2_\varphi \chi_{\text{u/d}} - \left| \kappa 
\right| \, \chi_{\text{d/u}} + \gamma\, \left| \chi_{\text{u/d}} \right|^2 \chi_{\text{u/d}} 
\quad .
\end{equation}
Here $\chi_{\text{u(d)}}$ is the condensate wave function for atoms in the upper (lower) ring.
Our equation (\ref{chiEqs}) agrees with Eq.~(2) of Ref.~\cite{vKlitz:prl:07}, except that we
explicitly indicate the negative sign of the tunnel coupling.  We give detailed reasons for the
relevance of the sign of the tunnel coupling below in Sec.~\ref{sec:sign}, where we also show
that the tunnel coupling is indeed negative. At this point, we only note that the tunnel coupling
$\kappa$ was assumed to be positive in Ref.~\cite{vKlitz:prl:07} (see their Fig.~1), in
contradiction to our findings. 

The most general form of the polar-angle-dependent wave function can be written as a
Fourier series, $\chi_{\text{u/d}} = (2\pi)^{-\frac{1}{2}} \sum_m \alpha_m^{\text{(u/d)}} \,
\ee^{i m \varphi}$. Inserting this \textit{Ansatz\/} into Eq.~(\ref{chiEqs}) and equating
coefficients of the orthogonal Fourier components, we find
\begin{equation}\label{alphaEqs}
i \partial_\tau \alpha_m^{\text{(u/d)}} = m^2 \alpha_m^{\text{(u/d)}} - |\kappa|\, 
\alpha_m^{\text{(d/u)}} +\frac{\gamma}{2\pi} \sum_{n, n^\prime}
\alpha_n^{\text{(u/d)}} \alpha_{n^\prime}^{\ast\text{(u/d)}} 
\alpha_{m-n+n^\prime}^{\text{(u/d)}} .
\end{equation}
This result differs from the corresponding Eq.~(3) in Ref.~\cite{vKlitz:prl:07} in the tunnel
coupling and the non-linear term.

As a first approximation, it is reasonable to assume that only the $m=0$ mode is occupied in
each of the two annuli. Straightforward calculation yields the new ground and excited state of
the coupled-annuli system, which are the symmetric and antisymmetric superpositions of
single-well states having chemical potential $\mu_\pm = \varepsilon \mp |\kappa|$,
respectively. $\varepsilon=\gamma N_0/(2\pi)$ is defined in terms of the equal number of
atoms $N_0$ in each well as in Ref.~\cite{vKlitz:prl:07}.
In order to study the stability of these states, finite but small amplitudes in the $m\ne 0$
modes are assumed:
\begin{equation}\label{perAnsatz}
\alpha_{m\ne 0}^{\text{(u/d)}} =\ee^{-i \mu_\pm \tau} \left[ u_{m,\pm}^{\text{(u/d)}}
\ee^{- i \omega \tau} + v_{m,\pm}^{\ast\text{(u/d)}} \ee^{i \omega \tau} \right] \quad .
\end{equation}
Here the subscript $\pm$ distinguishes perturbations to the ground and excited states,
respectively. Inserting the perturbation (\ref{perAnsatz}) into Eq.~(\ref{alphaEqs}) and
linearising in the small amplitudes $u, v$ yields
\begin{eqnarray}\label{BdGeqsFin}
\omega u_{m,\pm}^{\text{(u/d)}} &=& \left(m^2 +\varepsilon \pm |\kappa| \right)
u_{m,\pm}^{\text{(u/d)}} + \varepsilon v_{-m,\pm}^{\text{(u/d)}} - |\kappa|
u_{m,\pm}^{\text{(d/u)}} \, , \nonumber \\
- \omega v_{-m,\pm}^{\text{(u/d)}} &=& \left(m^2 + \varepsilon \pm |\kappa| \right)
v_{-m,\pm}^{\text{(u/d)}} + \varepsilon u_{m,\pm}^{\text{(u/d)}} - |\kappa|
v_{-m,\pm}^{\text{(d/u)}} \, . \nonumber \\
\end{eqnarray}
The upper (lower) sign refers to the symmetric ground (antisymmetric excited) state. Crucial
differences between our Eq.~(\ref{BdGeqsFin}) and Eq.~(5) in Ref.~\cite{vKlitz:prl:07} result
in markedly different excitation spectra. We find that both the symmetric (ground) state and
antisymmetric (excited) state share one branch,
\begin{subequations}
\begin{equation}
\omega_1 = \sqrt{\left( m^2 + \varepsilon \right)^2 - \varepsilon^2} \, .
\end{equation}
whose frequency is independent of the tunnel coupling. This was also found in
Ref.~\cite{vKlitz:prl:07}. In contrast to these authors, however, we find that the second branch 
differs for the two states:
\begin{equation} \label{eqn:unstable}
\omega_{2,\pm} = \sqrt{\left( m^2 + \varepsilon \pm 2 |\kappa| \right)^2 - 
\varepsilon^2} \, .
\end{equation}
\end{subequations}
Clearly, $\omega_{2,+}$ is always real for repulsive BECs ($\varepsilon>0$), implying stability
of the symmetric (ground) state of the coupled annular condensates. In contrast, the 
antisymmetric (excited) state will become unstable for $\varepsilon> |\kappa| - m^2/2 > 0$,
signified by $\omega_{2,-}$ becoming imaginary in this range. Our own numerical simulations
of the time evolution of the antisymmetric state seeded with a small amount of noise show the
development of angular-momentum Josephson junctions similar to those shown in Figs.~2
and 3 of Ref.~\cite{vKlitz:prl:07}.

In attractive condensates where $\varepsilon<0$, imaginary solutions of $\omega_1$ for
$2 \varepsilon < -m^2$ indicate the well-known modulational instability towards the formation
of localized peaks (bright solitons) in the individual rings. For the symmetric state,
$\omega_{2,+}$ does not add new instabilities (with imaginary solutions for $\varepsilon < -
m^2/2 - |\kappa|$). The antisymmetric state, however, is further destabilised by the tunnel
coupling due to imaginary frequencies of $\omega_{2,-}$ at $\varepsilon< |\kappa| - m^2/2
<0$.

\section{Sign of the tunnel coupling}
\label{sec:sign}

In our analysis so far we have assumed that the sign of the coupling constant $\kappa$ is 
negative. This lead to the symmetric state with  $\alpha_0^{\text{(d)}} = \alpha_0^{\text{(u)}}
= \text{const}\cdot e^{i\mu_+ \tau}$ and $\alpha_{m\ne 0}^{\text{(u/d)}}=0$ with $\mu_+ = 
\varepsilon - |\kappa|$ being the ground state. Let us now briefly consider the consequences
of the (hypothetical) case of a positive coupling  constant $\kappa>0$. The analysis of
Sec~\ref{sec:neg} can be carried out the same  way as before, with the difference that
$|\kappa|$ should be replaced by $-|\kappa|$ in all formulae. It is easily seen that, in this
case, the antisymmetric state with $\alpha_0^{\text{(d)}} =  -\alpha_0^{\text{(u)}}$ will be the
ground state. Since the sign change also affects Eq.~(\ref{eqn:unstable}), we find the
antisymmetric state being stable (for $\varepsilon>0$) and the symmetric one becoming
unstable. However, since the roles of these states have changed, we still find that the ground
state is stable for repulsive BECs.

\begin{figure}[b]
\centerline{\includegraphics[width=3in]{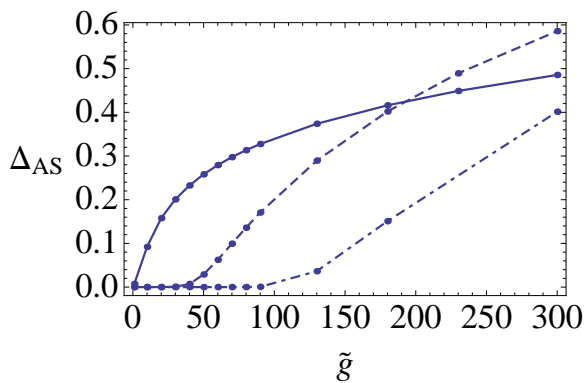}}
\caption{\label{fig:SAS}
Energy difference $\Delta_{\text{AS}}=E_{\text{A}}-E_{\text{S}}$ between the lowest
antisymmetric and symmetric eigenstates of a quadratic-plus-quartic double-well potential,
plotted as a function of the dimensionless effective interaction strength $\tilde g$. The fact
that $\Delta_{\text{AS}}\ge 0$ indicates that the symmetric (node-less) state remains the
ground state even in the limit where the atoms interact strongly. Double-well parameters
[see Eq.~(\ref{eq:dwell})] are $\xi_0=5$ and $h=0.002$ (solid curve), 0.02 (dashed curve),
0.05 (dot-dashed curve).}
\end{figure}

\begin{figure*}[t]
\centerline{\includegraphics[width=3in]{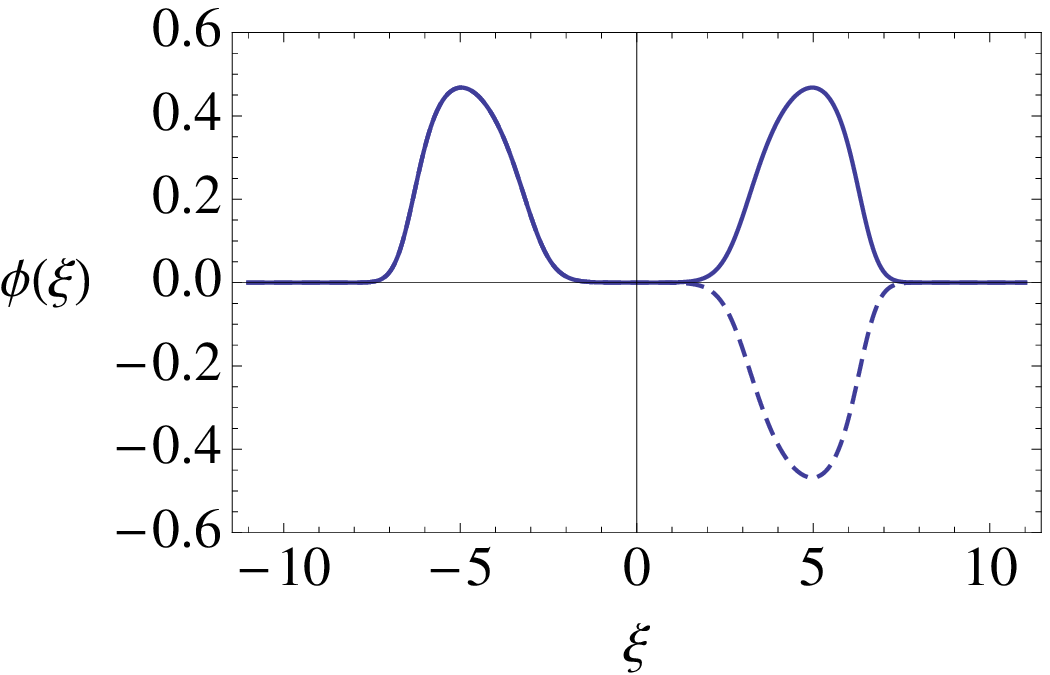}\hspace{1cm}
\includegraphics[width=3in]{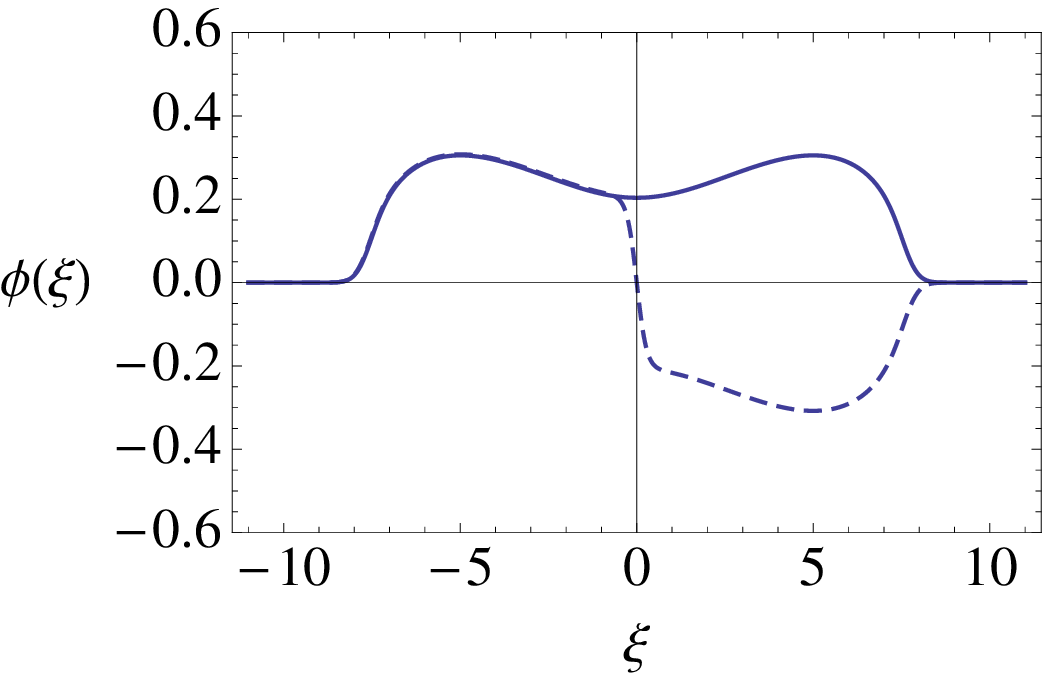}}
\caption{\label{fig:wavefunct}
Lowest symmetric (solid curve) and antisymmetric (dashed curve) condensate wavefunctions
obtained for a double-well potential [see Eq.~(\ref{eq:dwell})] with $\xi_0=5$, $h=0.05$, and
$\tilde g = 30$ (left panel) or 300 (right panel). Notice the greater delocalization of atoms
between the two wells when the interaction strength is high. This arises because repulsive
interactions result in an effective lowering of the tunnel barrier.}
\end{figure*}

In order to determine the correct sign and value of the coupling constant $\kappa$ appearing
in Eq.~(\ref{chiEqs}), we briefly revisit the derivation of this model. For the purpose of finding $\kappa$, the azimuthal degree of freedom in the double-ring model of Ref.~\cite{vKlitz:prl:07} is irrelevant and it suffices to consider the problem of a BEC in a one-dimensional (1D) double-well potential, as in Refs.~\cite{milburn:pra:97,ananikian:pra:06}.
Generalisation to multiple wells and different geometries (coupled cigars or pancakes) are
straightforward.

Different derivations of effective two-mode models have been presented in the literature \cite{milburn:pra:97,ananikian:pra:06,javanainen,pitaevskii}. 
The goal of a two-mode model is generally to correctly describe the ground and low-lying excited states of the system.
The quantity that is obtainable from the 1D model and carries unambiguous information
about the sign of the tunnel coupling is the energy difference $\Delta_{\text{AS}} = E_{\text{A}}
- E_{\text{S}}$ between the antisymmetric state with one node and the node-less symmetric
state. 
In the simplest case, the tunnel coupling $\kappa$ is determined from the single-particle linear Schr\"odinger equation.
This approach is commonly used when deriving the fully quantum-mechanical Bose-Hubbard model \cite{javanainen,fisher} and was the basis of Ref. \cite{milburn:pra:97}.
In this case both the sign and the value of $\kappa$ are completely independent of particle number or interaction strength. The node
theorem of quantum mechanics~\cite{messiah_node} guarantees that the node-less
symmetric state in a one-dimensional double-well potential must be the ground state, thus
$\Delta_{\text{AS}}\ge 0$ and consequently, the correct sign of  $\kappa$ is negative.

In a more general class of models based on mean-field theory, the parameters of the two-mode model are chosen in order to reproduce $\Delta_{\text{AS}}$ as found from a one-dimensional GP equation. The ordering of eigenvalues of the GP equation by the number of nodes in the wave function is now no longer guaranteed by the node theorem of linear quantum mechanics and we are not aware of a non-linear generalization of this theorem. However, we find by numerical calculation that the ordering is preserved under repulsive interactions.
The main result of this section is  the dependence of $\Delta_{\text{AS}}$ on the non-linear interaction
strength $\tilde g$, shown in Fig.~\ref{fig:SAS}. As can be seen from Fig.~\ref{fig:SAS}, the presence of a repulsive
non-linear interaction does not change the sign of $\Delta_{\text{AS}}$ and therefore $\kappa$ remains negative. We now present details of our calculation.

Starting from the three-dimensional GP equation for a BEC in a double-well or double-ring trap and
employing a separation ansatz, an effective 1D equation describing the dynamics in the
direction perpendicular to the potential barrier can be derived:
\begin{equation}\label{eq:1DGPE}
\frac{\mu}{\varepsilon_0}\, \phi(\xi) = \left[ -\frac{d^2}{d\xi^2} + V_{\text{dw}}(\xi) + \tilde g
\left| \phi(\xi) \right|^2 \right] \phi(\xi) \, .
\end{equation}
Here the energy scale $\varepsilon_0$ and length scale $a_0$ defined by the trap are used
as units for all energies and the spatial coordinate, respectively, and the condensate wave
function $\phi$ is normalized to unity. We introduced the dimensionless interaction strength
$\tilde g = g_{\text{1D}}N/(\varepsilon_0 a_0)$, where $N$ denotes the number of atoms in
the trap and $g_{\text{1D}}$ is the effective 1D interaction strength~\cite{olshani:prl:98}. To
be specific, we use the double-well potential
\begin{equation}\label{eq:dwell}
V_{\text{dw}} = h \left( \xi^2 - \xi_0^2\right)^2 \quad ,
\end{equation}
where $h$ parameterizes the barrier height between the two wells centered at $\pm\xi_0$.
It is straightforward to solve Eq.~(\ref{eq:1DGPE}) with the potential (\ref{eq:dwell}) and find
the lowest symmetric and antisymmetric eigenstates as well as their respective energies
$E_{\text{S}}$ and $E_{\text{A}}$. Figure~\ref{fig:wavefunct} shows typical results obtained
for low and high interactions strengths, respectively. As is apparent from the figure, the
higher repulsive interaction strength is associated with more strongly delocalized
double-well wave functions, indicating an effectively stronger tunnel coupling. This can be
explained simply by noting that the nonlinear interaction energy for the two condensate
fractions in each well shifts up their respective energies, thus effectively lowers the barrier
and brings the condensates closer together. As a result, the effective tunnel coupling
increases. Most importantly, the energy difference between the lowest symmetric and
antisymmetric eigenstates remains positive for any strength of repulsive interactions, which
implies that the sign of the tunnel coupling $\kappa$ entering Eq.~(\ref{chiEqs}) is negative.

\section{Conclusions}
\label{sec:concl}

Reference \cite{vKlitz:prl:07} predicts a dynamical instability of a repulsively interacting BEC in a double-ring trap against angular momentum fluctuations. We have carefully revisited the analysis of Ref.~\cite{vKlitz:prl:07} and have recalculated the elementary excitation spectrum. This leads us to a different conclusion that makes physical sense.
The ground state of a non-rotating condensate in the double-ring configuration is stable
against spontaneous angular momentum oscillations. 
However, the antisymmetric state with
its circular node between the two annular quantum wells can be viewed as the analog of a
stationary 2D dark soliton, which is known to have a dynamical instability towards the
formation of local vorticity ("snake'' instability)~\cite{brand:pra:02}.

\acknowledgments

JB is supported by the Marsden Fund Council (contract MAU0706) from Government
funding, administered by the Royal Society of New Zealand.


\end{document}